\documentclass[aps,prd,twocolumn,superscriptaddress,showpacs,amsfonts,amssymb,amsmath]{revtex4}
\usepackage[dvips]{graphicx}
\DeclareGraphicsExtensions{.ps,.eps,.pdf}
 \DeclareGraphicsExtensions{.jpg,.png}
\usepackage{color}
\usepackage{ulem}
\usepackage[colorlinks=true,linkcolor=blue,citecolor=blue,urlcolor=blue]{hyperref}
\usepackage{tabularx}
\usepackage{multirow}

\begin{document}

\title{Magnetic properties of the honeycomb oxide Na$_2$Co$_2$TeO$_6$}

\author{E. Lefran\c{c}ois}
\email[]{lefrancois@ill.fr}
\affiliation{Institut N\'eel, CNRS, 38042 Grenoble, France}
\affiliation{Universit\'e Grenoble Alpes, 38042 Grenoble, France}
\affiliation{Institut Laue-Langevin, 38042 Grenoble, France}
\author{M. Songvilay}
\affiliation{Institut N\'eel, CNRS, 38042 Grenoble, France}
\affiliation{Universit\'e Grenoble Alpes, 38042 Grenoble, France}
\affiliation{CEA, Centre de Saclay, /DSM/IRAMIS/Laboratoire L\'eon Brillouin, 91191 Gif-sur-Yvette,France}
\author{J. Robert}
\affiliation{Institut N\'eel, CNRS, 38042 Grenoble, France}
\affiliation{Universit\'e Grenoble Alpes, 38042 Grenoble, France}
\author{G. Nataf}
\affiliation{Institut N\'eel, CNRS, 38042 Grenoble, France}
\affiliation{Universit\'e Grenoble Alpes, 38042 Grenoble, France}
\affiliation{Materials Research and Technology Department, Luxembourg Institute of Science and Technology, 
41 rue du Brill, L-4422 Belvaux, Luxembourg}
\affiliation{SPEC, CEA, CNRS, Université Paris-Saclay, CEA Saclay, 91191 Gif-sur-Yvette Cedex, France}
\author{E. Jordan}
\affiliation{Institut N\'eel, CNRS, 38042 Grenoble, France}
\affiliation{Universit\'e Grenoble Alpes, 38042 Grenoble, France}
\affiliation{IMEP-LAHC, F-38000 Grenoble, France}
\author{L. Chaix}
\affiliation{Institut N\'eel, CNRS, 38042 Grenoble, France}
\affiliation{Universit\'e Grenoble Alpes, 38042 Grenoble, France}
\affiliation{Stanford Institute for Materials and Energy Sciences, SLAC National Accelerator Laboratory, Menlo Park, California 94025, USA}
\author{C. V. Colin}
\affiliation{Institut N\'eel, CNRS, 38042 Grenoble, France}
\affiliation{Universit\'e Grenoble Alpes, 38042 Grenoble, France}
\author{P. Lejay}
\affiliation{Institut N\'eel, CNRS, 38042 Grenoble, France}
\affiliation{Universit\'e Grenoble Alpes, 38042 Grenoble, France}
\author{A. Hadj-Azzem}
\affiliation{Institut N\'eel, CNRS, 38042 Grenoble, France}
\affiliation{Universit\'e Grenoble Alpes, 38042 Grenoble, France}
\author{R. Ballou}
\affiliation{Institut N\'eel, CNRS, 38042 Grenoble, France}
\affiliation{Universit\'e Grenoble Alpes, 38042 Grenoble, France}
\author{V. Simonet}
\email[]{virginie.simonet@neel.cnrs.fr}
\affiliation{Institut N\'eel, CNRS, 38042 Grenoble, France}
\affiliation{Universit\'e Grenoble Alpes, 38042 Grenoble, France}

\date{\today}

\begin{abstract}
We have studied the magnetic properties of Na$_2$Co$_2$TeO$_6$, which features a honeycomb lattice of magnetic Co$^{2+}$ ions, through macroscopic characterization and neutron diffraction on a powder sample. We have shown that this material orders in a zig-zag antiferromagnetic structure. In addition to allowing a linear magnetoelectric coupling, this magnetic arrangement displays very peculiar spatial magnetic correlations, larger in the honeycomb planes than between the planes, which do not evolve with the temperature. We have investigated this behavior by classical Monte Carlo calculations using the $J_1$-$J_2$-$J_3$ model on a honeycomb lattice with a small interplane interaction. Our model reproduces the experimental neutron structure factor, although its absence of temperature evolution must be due to additional ingredients, such as chemical disorder or quantum fluctuations enhanced by the proximity to a phase boundary.
\end{abstract}


\maketitle


\section{Introduction}

\begin{figure}
\resizebox{8.6cm}{!}{\includegraphics{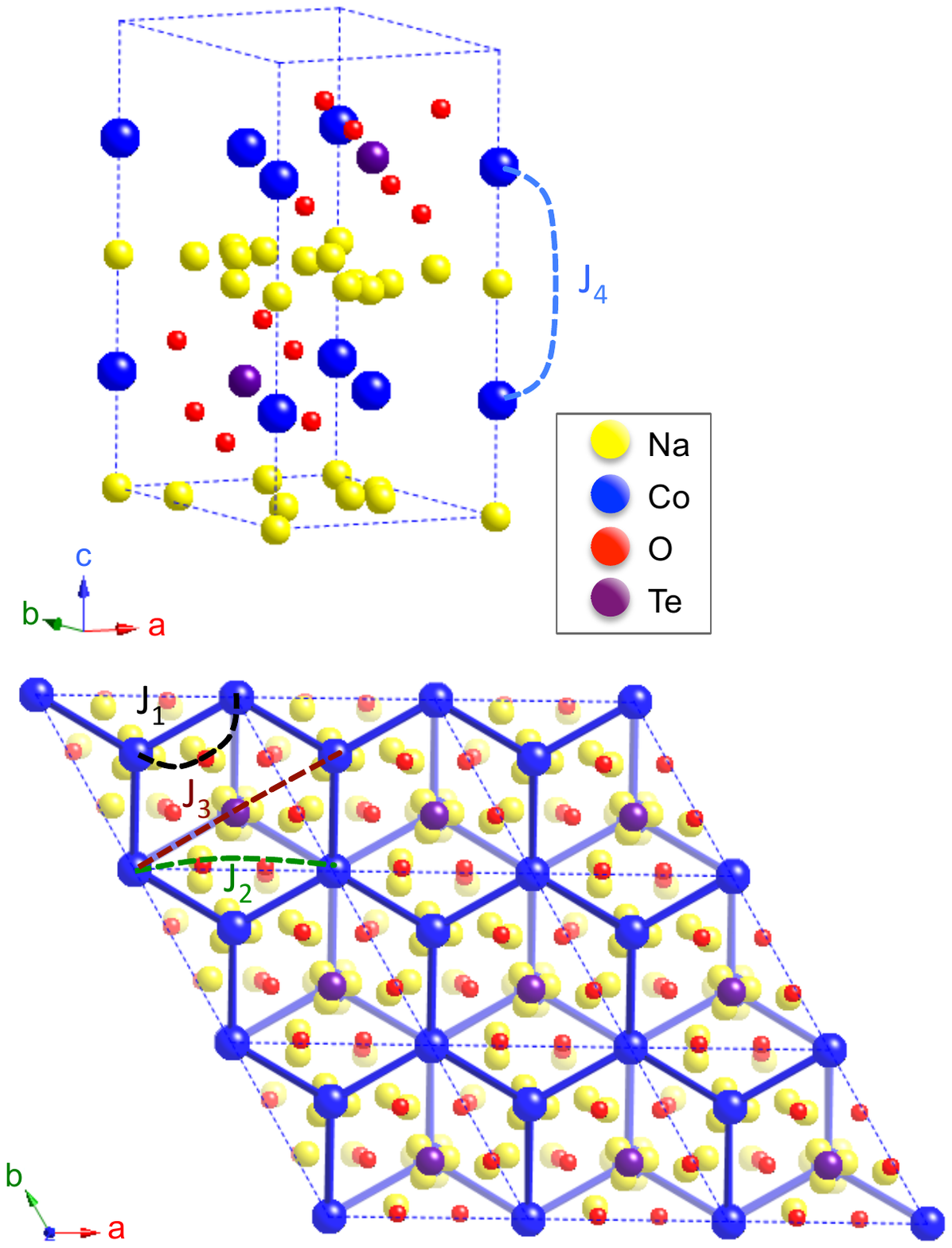}}
\caption{Crystal structure of Na$_2$Co$_2$TeO$_6$, in a perspective view (top) and projected along the $c$-axis (bottom) with the two Co$^{2+}$ honeycomb layers underlined in blue. The Na$^+$ ion multiple positions, too close to be simultaneously fully occupied, are represented with yellow spheres, highlighting the Na disorder. The main magnetic couplings, in-plane $J_1$ to $J_3$ and out-of-plane $J_4$, are shown by dashed lines.}
\label{fig1}
\end{figure}

There is a longstanding interest in the potentially unconventional electronic and magnetic properties of honeycomb lattices of magnetic atoms. Most recently, the realization of a new kind of spin liquid predicted by Kitaev was searched in real materials with 4d or 5d electrons \cite{Kitaev2006}. This state of matter is achieved in presence of anisotropic (bond-directional) interactions favored by a strong spin-orbit coupling \cite{Jackeli2009}. On the other hand, in the isotropic Heisenberg model, manifestations of magnetic frustration are also expected in line with the exotic phase diagram obtained when magnetic interactions beyond the first neighbors are present \cite{Rastelli1979,Fouet2001,Albuquerque2011,Reuther2011,Oitmaa2011,Li2016a,Li2016b,Rehn2015}. In addition to conventional ordered phases, these lead to ground states degeneracies, possibly lifted by the order-by-disorder mechanism, or favoring nonmagnetic phases in the quantum (e.g. valence bond crystal or spin liquid) or classical (e.g. classical spin liquid) regimes. Another theoretical approach, based on the spin-1/2 Hubbard model at half-filling, in the intermediate coupling regime in the vicinity of the Mott transition, also discloses a quantum spin liquid state based on resonating valence-bonds \cite{Meng2010}. In this context, new materials that might display these potential exotic behaviors are intensively looked for. Viciu {\it et al.} have recently reported the synthesis of two new oxides, Na$_2$Co$_2$TeO$_6$ and Na$_3$Co$_2$SbO$_6$, with a perfect honeycomb lattice of magnetic Co$^{2+}$ ions \cite{Viciu2007}. Actually, these materials belong to large families of compounds allowing various substitutions, in particular Na-Li, Sb-Bi and Cu-Co-Ni for the magnetic ions \cite{Skakle1997,Zvereva2012,Zvereva2015a,Zvereva2015b,Miura2006,Schmidt2013,Berthelot2012a,Sankar2014,Xu2005,Berthelot2012b,Seibel2013,Wong2016}. In this article, we will concentrate on Na$_2$Co$_2$TeO$_6$ for which no extensive investigation has been reported yet beyond the work of Viciu {\it et al.} \cite{Viciu2007}. In addition to the above quest for exotic magnetic behaviors, Na$_2$Co$_2$TeO$_6$ is structurally related to the Na$_x$CoO$_2$ cobaltates, that exhibit in particular superconductivity through water molecules intercalation \cite{Takada2003}. Moreover, the space group of Na$_2$Co$_2$TeO$_6$ is non-centrosymmetric which is a necessary condition for ferroelectricity. It is thus expected that, depending on the stabilized magnetic order, this compound could display interesting multiferroic/magnetoelectric properties. 

We hereafter report the magnetic properties of Na$_2$Co$_2$TeO$_6$ probed by magnetization, specific heat and neutron scattering measurements. We unveil the nature of the low temperature magnetic phase, whose incomplete long-range ordering is discussed using Monte-Carlo calculations.


\section{Synthesis, structure and experimental details}
A polycrystalline sample of Na$_2$Co$_2$TeO$_6$ was prepared by mixing Na$_2$CO$_3$ and Co$_3$O$_4$ with TeO$_2$, with a 8 days thermal treatment at 800$^{\circ}$C under argon atmosphere with intermediate homogenization, as described in Ref. \onlinecite{Viciu2007}. The structure and quality of the sample were checked by x-ray diffraction. Na$_2$Co$_2$TeO$_6$ presents a two-layer hexagonal structure, which can be described by the non-centrosymmetric space group P6$_3$22 (No 182). The Co$^{2+}$ ions are in an octahedral environment and occupy the 2$b$ Wyckoff site with atoms at (0, 0, 1/4) and (0, 0, 3/4) and the 2$d$ Wyckoff site with atoms at (1/3, 2/3, 1/4) and (2/3, 1/3, 3/4). They are arranged on a perfect honeycomb lattice. The Na distribution between the honeycomb layers is on the other hand highly disordered and site distributed (see Fig.~\ref{fig1}). 

\begin{figure}
\resizebox{8.6cm}{!}{\includegraphics{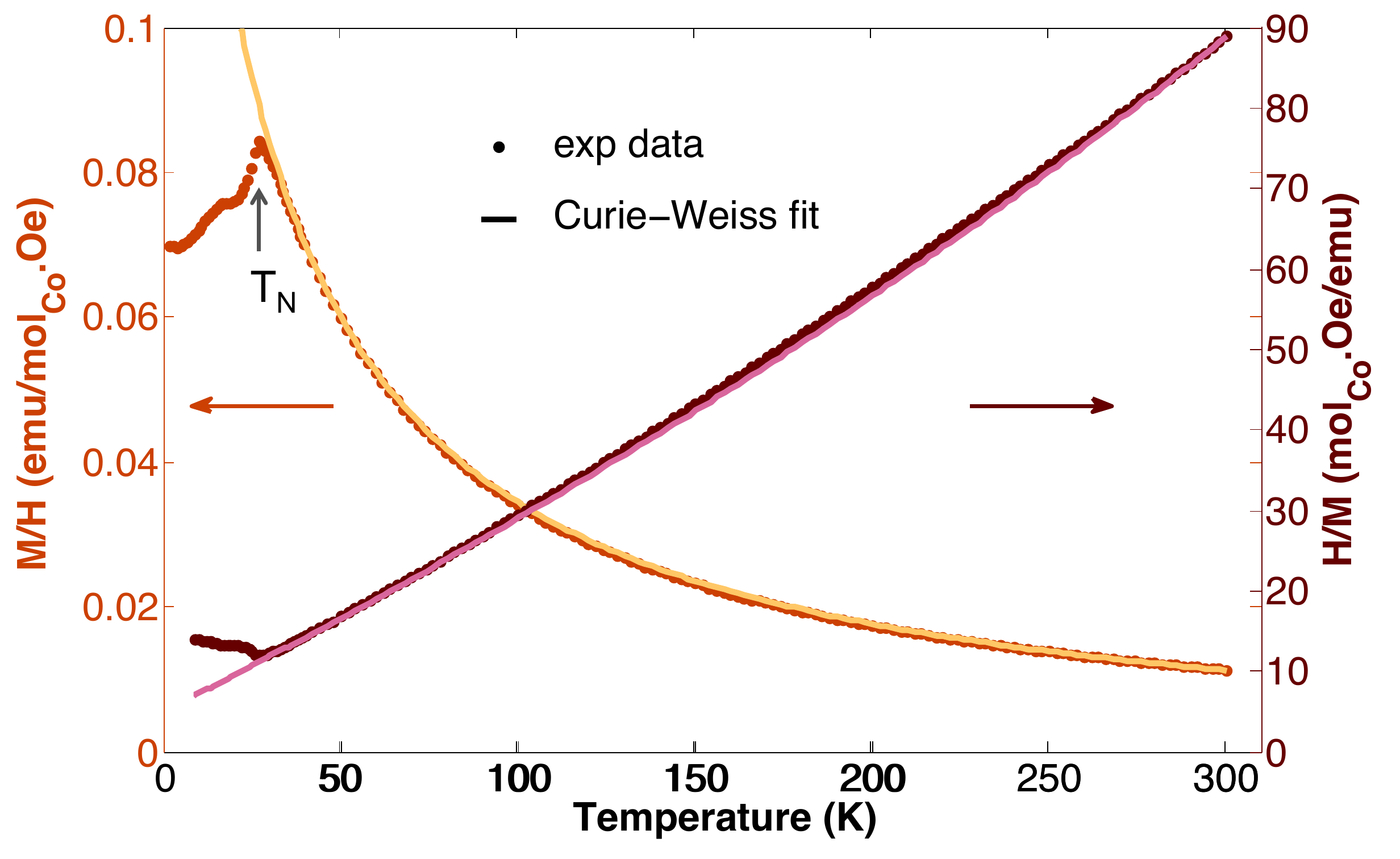}}
\caption{Temperature dependence of the susceptibility $M/H$ measured with a 1~T magnetic field and of the inverse susceptibility. The Curie-Weiss fit (plain line) was obtained in the 100-300~K range for the parameters $\chi_0$=-0.0029(2)~emu/mol$_{\rm Co}$.Oe, $C$=4.55(4)~emu.K/mol$_{\rm Co}$.Oe and $\theta$=-23.3(1)~K.}
\label{fig2}
\end{figure}

\begin{figure}
\resizebox{8.6cm}{!}{\includegraphics{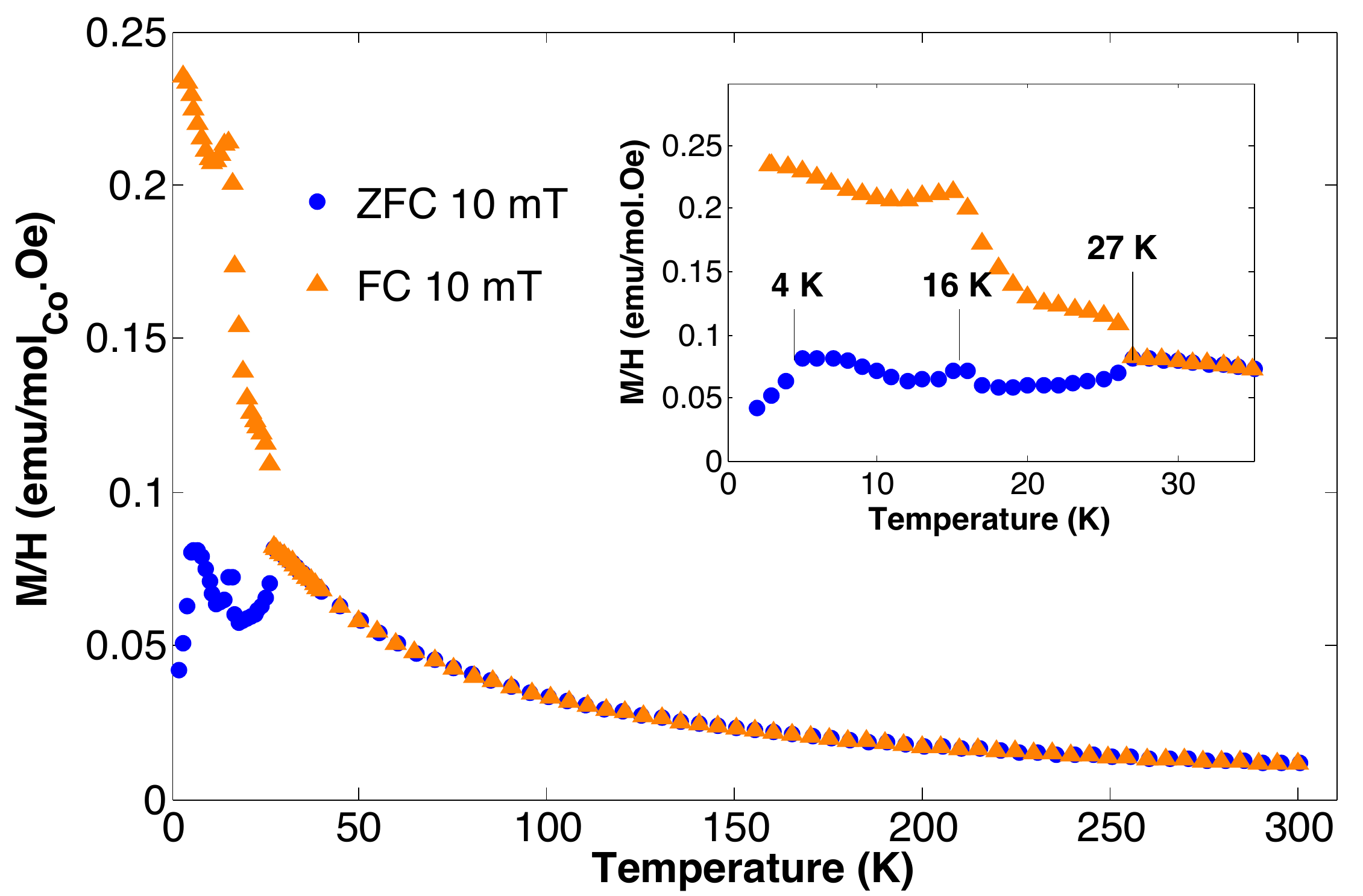}}
\caption{Field-cooled and zero-field-cooled measurements of the magnetization versus temperature in a magnetic field of 0.01~T with a zoom of the transition temperature range in the inset.}
\label{fig3}
\end{figure}

The magnetic properties of the Na$_2$Co$_2$TeO$_6$ powder sample were investigated under magnetic fields up to 5~T in the temperature range from 2 to 300~K with a Quantum Design MPMS\circledR\ superconducting quantum interference device magnetometer and up to 10.5~T from 2 to 300~K with a purpose-built extraction magnetometer. The specific heat of the sample was measured with a Quantum Design PPMS\circledR\ relaxation-time calorimeter by increasing the temperature from 2 to 300~K.

Neutron powder-diffraction measurements were performed at the Institut Laue-Langevin using the high resolution two-axis powder diffractometer CRG-D1B (wavelength $\lambda$=2.52 \AA). The thermal evolution of the diffraction patterns was recorded using an orange cryostat by decreasing the temperature from 300 to 2~K. Long scans at 2~K, 12~K and at 35~K (below and above $T_N$=27~K) were performed. 

\begin{figure}
\resizebox{8.6cm}{!}{\includegraphics{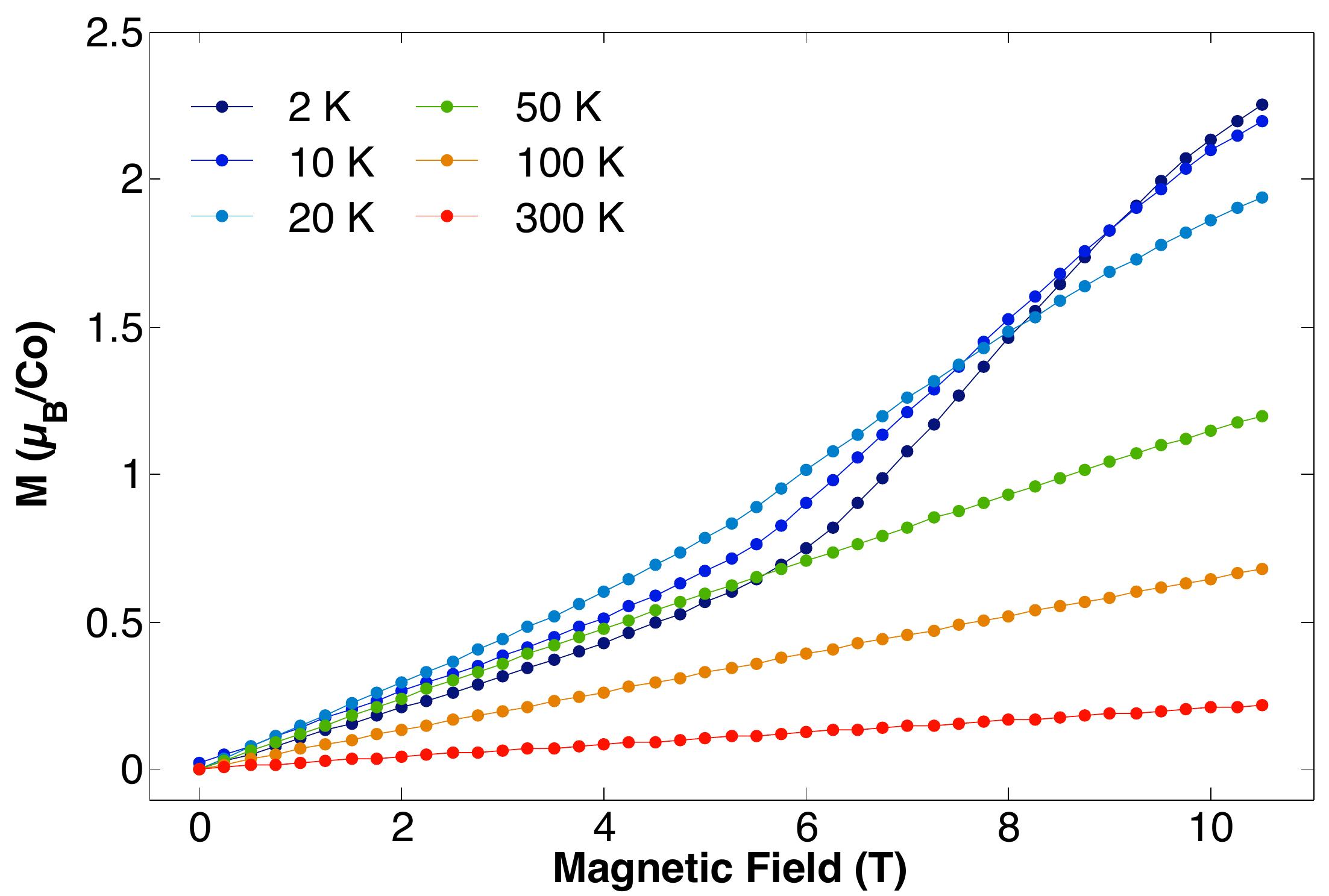}}
\caption{Magnetic field dependence of the magnetization of Na$_2$Co$_2$TeO$_6$ at different temperatures. }
\label{fig4}
\end{figure}

\section{Macroscopic characterization}

The temperature and field-dependent magnetization $M$ of Na$_2$Co$_2$TeO$_6$ are shown in Figs.~\ref{fig2}, \ref{fig3} and \ref{fig4}. A cusp in the susceptibility ($\chi=M/H$ in the linear regime) measured in a field of 1~T is observed at $T_N$=27~K, consistent with a transition from a paramagnetic state toward an antiferromagnetic (AF) order. Additional features are visible near 16 and 4~K respectively. These transitions are also visible in the field-cooled/zero-field-cooled measurements performed in a small field of 0.01 T (see Fig.~\ref{fig3}). In particular, a marked bifurcation between the two measurements, that signals thermomagnetic irreversibilities, is observed below $T_N$. A Curie-Weiss fit, $\chi=\chi_0+C/(T-\theta)$ including a diamagnetic contribution $\chi_0$, was performed in the [100-300~K] temperature range, following Viciu e{\it t al.} \cite{Viciu2007} who pointed out that this diamagnetic contribution is responsible for the curvature of the inverse $M/H$ up to high temperature. This fit yields the effective moment $\mu_{\rm eff}$=5.64~$\mu_B$ which is within the range of calculated values for Co$^{2+}$ ions in a spin 3d$^7$ configuration with a non-zero orbital contribution. The Curie-Weiss temperature $\theta$=-23.3 K indicates dominant antiferromagnetic interactions in first approximation. As shown in Fig.~\ref{fig4}, the magnetic isotherms $M(H)$ deviate from linearity below 50 K due to the beginning of the magnetization saturation. Below 20~K, i.e. in the ordered phase, they present an upward curvature around 6 T signaling a possible metamagnetic transition. Our results are globally consistent with previous studies \cite{Viciu2007,Berthelot2012a}. In these works, the $T_N$ was reported respectively by Viciu {\it et al.} and Berthelot {\it et al.} at 17.7~K and 26~K and a second anomaly in the susceptibility was reported at 9~K and around 17~K (the 4 K anomaly observed in the present work was not mentioned). This variability in the transition temperatures suggests some sample dependency possibly associated to the Na disorder or due to some H$_2$O intercalation, already reported in alkali-metal cobaltates. 

\begin{figure}
\resizebox{8.6cm}{!}{\includegraphics{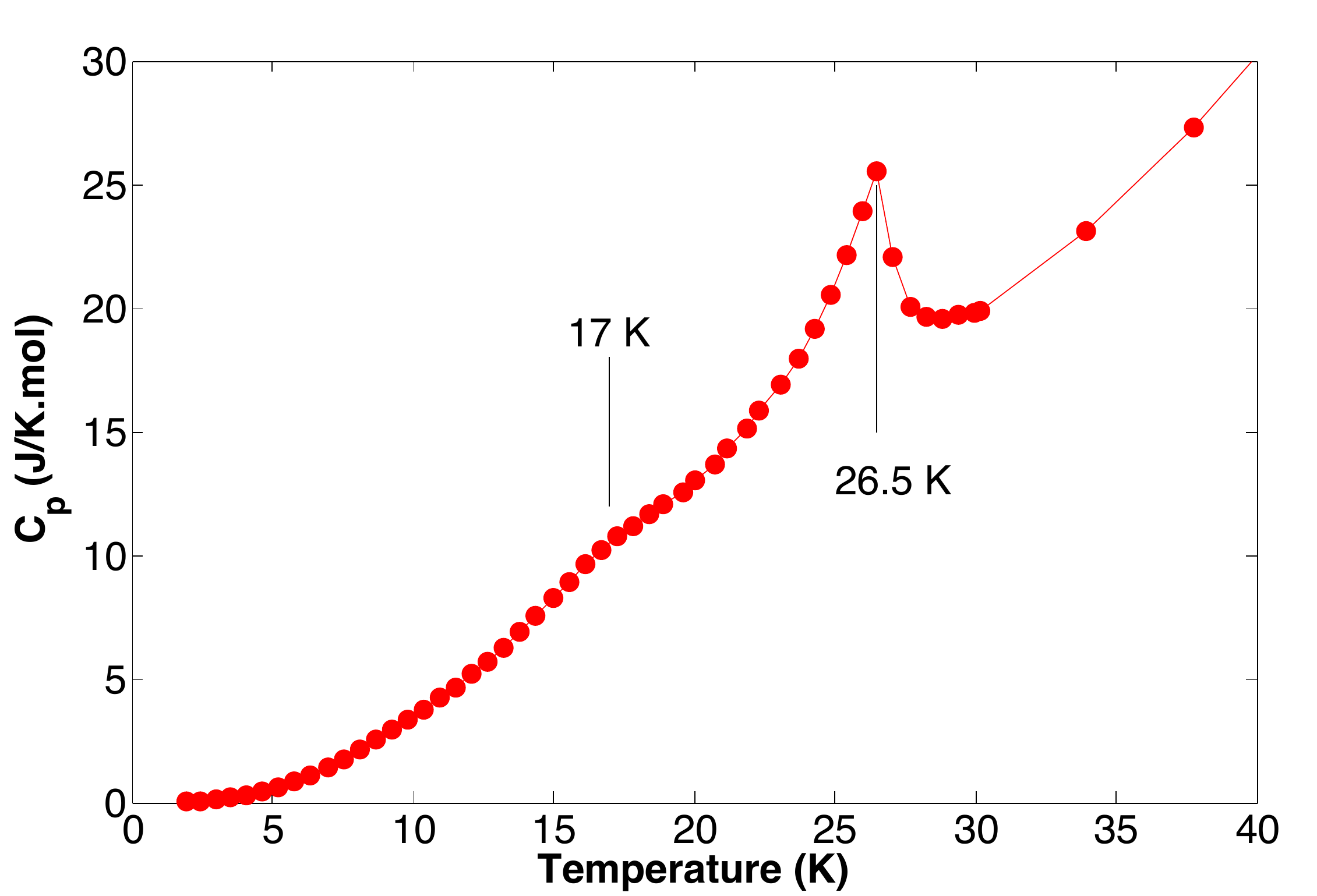}}
\caption{Temperature dependence of the specific heat corrected from the signal of the addenda.}
\label{fig5}
\end{figure}

Specific heat $C_p$ measurements on Na$_2$Co$_2$TeO$_6$ confirm the results from magnetization (see Fig.~\ref{fig5}): A sharp transition is observed at 26.5~K consistent with the N\'eel temperature. A broad shoulder is also visible at 17~K in agreement with the feature observed in the magnetization below T$_N$ and which is the possible signature of a spin reorientation.

\begin{figure}
\resizebox{8.6cm}{!}{\includegraphics{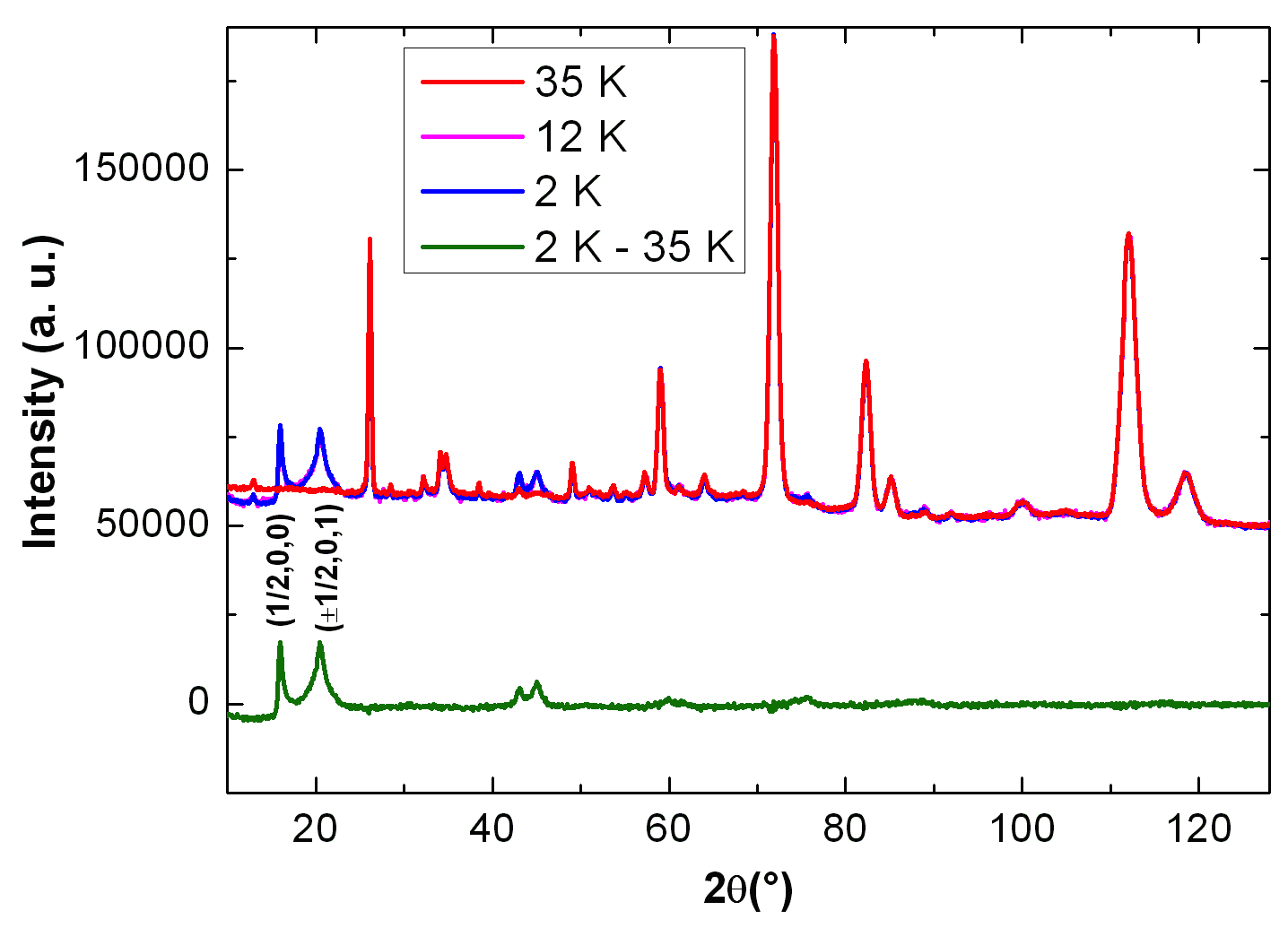}}
\caption{Powder neutron diffractograms at 35~K (red), 12~K (pink) and 2~K (blue) with the 2~K - 35~K difference (green). The 2~K and 12~K diffractograms are almost identical. The indexation of the first two magnetic Bragg reflections is given.}
\label{fig6}
\end{figure}


\section{Neutron diffraction}

In order to determine the magnetic order stabilized below $T_N$, powder neutron diffraction was performed. The nuclear structure was refined using the FullProf Suite \cite{Rodriguez1993} and found in agreement with the published one (see Table \ref{TableStruct}) \cite{Viciu2007} with a tiny amount of parasitic phase. The Na$^+$ ions partially occupy three sites with very little change in their occupation between 300 and 35~K. Slight shifts of the Na position on the two 12$i$ Wyckoff sites are found between these two temperatures.

\begin{table}
 \caption{Structural refinement of Na$_2$Co$_2$TeO$_6$ in the P6$_3$22 space group at 300~K with $a$=5.2694(5) \AA\ and $c$=11.231(3) \AA\ and a RF-factor= 9.95 (top), and at 35 K with $a$=5.2627(4) \AA\ and $c$=11.182(2) \AA\ and a RF-factor=9.25 (bottom).}
 \label{tab:Ireps}
 \begin{ruledtabular}
 \begin{tabular}{cccccc}
 Atom &Wyckoff & $x$ & $y$ & $z$ & Occ. \\
  \hline
 Co(1) & 2$b$ & 0 & 0 & 1/4 & 1 \\
 Co(2) & 2$d$ & 2/3 & $1/3$ & 1/4 & 1 \\
 Te & 2$c$ & 1/3 & 2/3 & 1/4 & 1 \\
 O & 12$i$ & 0.656(2) & -0.018(1) & 0.3459(3) & 1\\
 Na(1) & 12$i$ & 0.15(2) & 0.57(2) & -0.028(8) & 0.097(7)\\
 Na(2) & 2$a$ & 0 & 0 & 0 & 0.13(2)\\
 Na(3) & 12$i$ & 0.65(1) & 0.04(1) & 0.099(9) & 0.242(4)\\
  \hline
   \hline
 Co(1) & 2$b$ & 0 & 0 & 1/4 & 1 \\
 Co(2) & 2$d$ & 2/3 & $1/3$ & 1/4 & 1 \\
 Te & 2$c$ & 1/3 & 2/3 & 1/4 & 1 \\
 O & 12$i$ & 0.658(2) & -0.017(1) & 0.3469(3) & 1\\
 Na(1) & 12$i$ & 0.19(1) & 0.62(2) & -0.03(1) & 0.082(7)\\
 Na(2) & 2$a$ & 0 & 0 & 0 & 0.13(2)\\
 Na(3) & 12$i$ & 0.67(1) & 0.05(1) & -0.010(7) & 0.240(3)\\
 \end{tabular}
 \label{TableStruct}
 \end{ruledtabular}
\end{table}  

Below $T_N$=27~K, additional Bragg peaks characteristics of the magnetic order appear, which can be indexed by the propagation vector (1/2, 0, 0) (see Fig. \ref{fig6}). This corresponds to an antiferromagnetic arrangement with a magnetic cell doubled along $a$ with respect to the crystallographic cell. The difference between the curves measured at 2~K and at 35~K, isolating the magnetic signal, presents peculiar features: first a step-like increase of the signal is visible around 2$\theta$=15.5$^{\circ}$, coinciding with the rise of the first magnetic reflection. Then the shape of the peaks is not identical for all the reflections: some are narrow like the first one at 15.9$^{\circ}$, or much broader and with a lorentzian shape like the second one at 20.5$^{\circ}$. They correspond respectively to the (1/2, 0, 0) reflection and to the (1/2, 0, 1) and (-1/2, 0, 1) reflections. The first one is uniquely sensitive to the magnetic correlations along the $a$ direction, i.e. within the honeycomb planes, whereas the others are also sensitive to the out-of-plane correlations. The broader second peak might then be an indication of shorter-range correlations in the $c$ direction. The diffractograms are almost superimposed between 2 and 12~K (see Fig. \ref{fig6}). Note also that we have not observed any marked change in the diffractograms below $T_N$, in particular around 17~K where the second anomaly is seen in macroscopic measurements.

\begin{table}
 \caption{The symmetry operators associated with the four possible irreducible representations of the magnetic structure, based on the application of a (1/2, 0, 0) propagation vector.}
 \label{tab:Ireps}
 \begin{ruledtabular}
 \begin{tabular}{ccccc}
 &Sym. 1 & Sym. 2 & Sym. 3 & Sym. 4 \\
 &1: 0,0,0 & 2: (0,0,1/2) 0,0,z & 2: 0,y,0 & 2: 2x,x,1/4 \\
 \hline
 IRep(1) & 1 & 1 & 1 & 1 \\
 IRep(2) & 1 & 1 & $-1$ & $-1$ \\
 IRep(3) & 1 & $-1$ & 1 & $-1$ \\
 IRep(4) & 1 & $-1$ & $-1$ & 1\\
 \end{tabular}
 \label{TableIrreps}
 \end{ruledtabular}
\end{table}  

\begin{figure}
\resizebox{8.6cm}{!}{\includegraphics{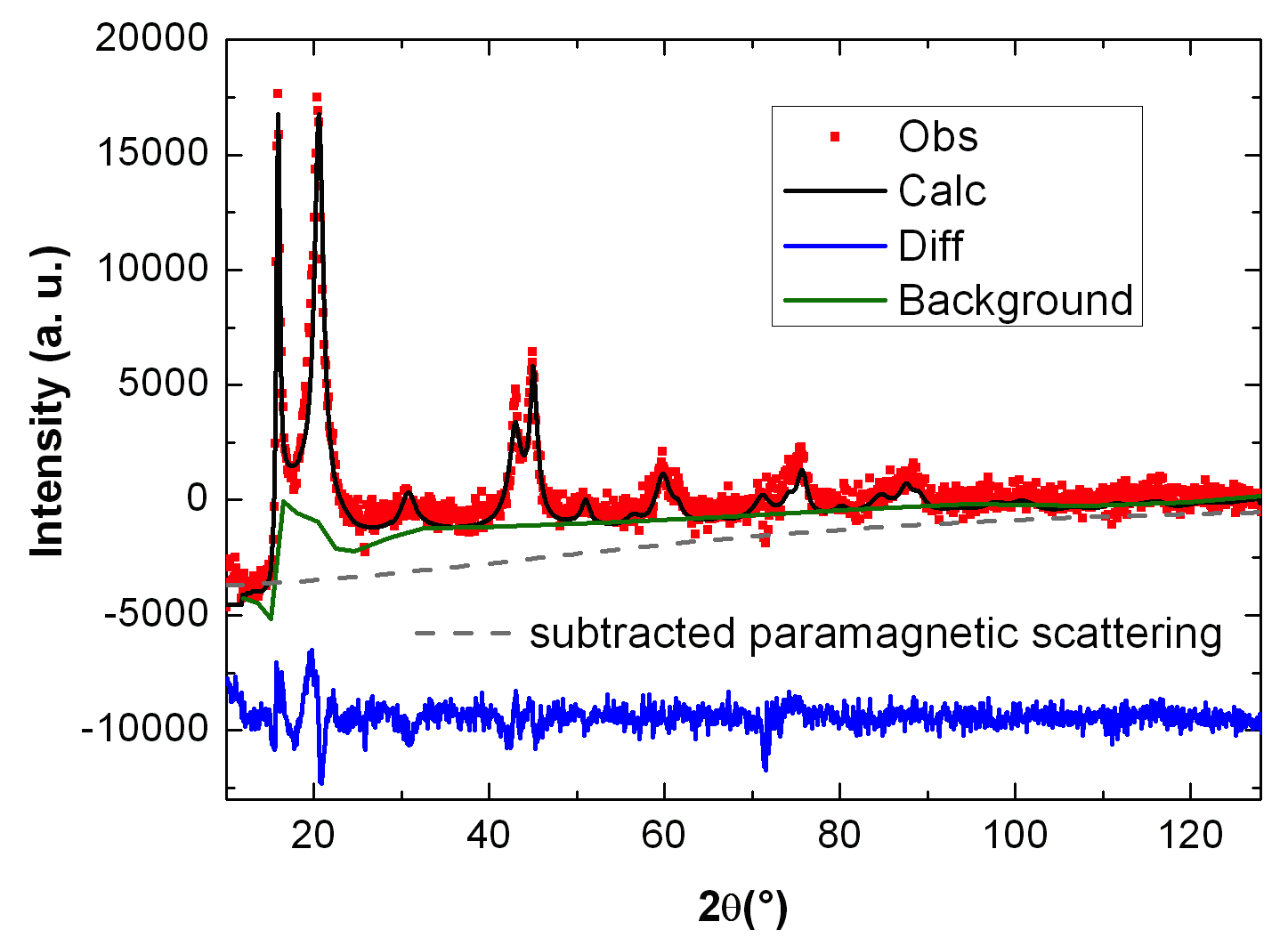}}
\caption{Rietveld refinement, using the IRep(2), of the difference between the diffractograms at 2~K and at 35~K. The measurements are in red, the calculation in black and the difference between the two in blue. The background used for the refinement is shown in green. The agreement factors of the fit are RF-factor=6.736 and global $\chi^2$=0.925. The grey dashed line is the assumed subtracted paramagnetic scattering that has been restored in figures \ref{fig12} and \ref{fig13}. }
\label{fig7}
\end{figure}

\begin{figure}
\resizebox{8cm}{!}{\includegraphics{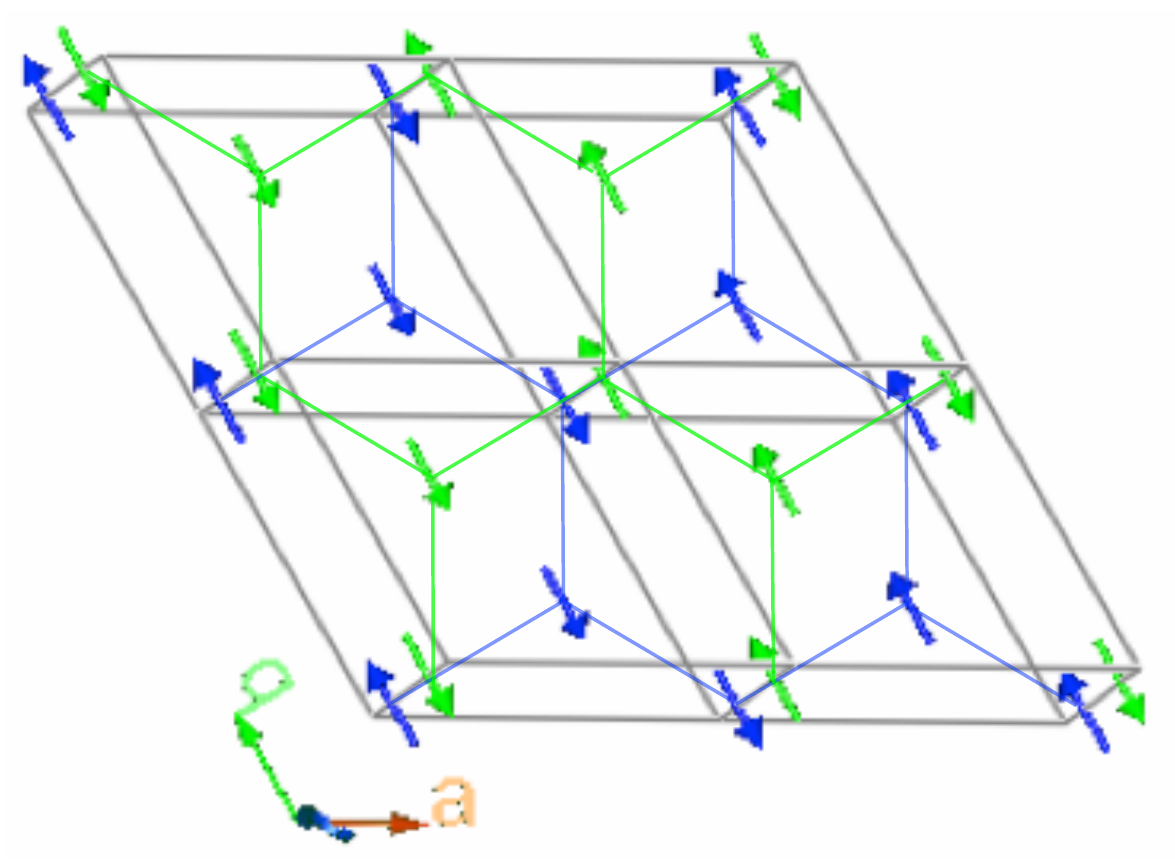}}
\caption{Zig-zag magnetic structure of Na$_2$Co$_2$TeO$_6$ in agreement with the podwer neutron diffractograms. The two colors depict the magnetic moments in the honeycomb layers at $z=0.25$ (blue) and at $z=0.75$ (green). The two sites, 2$b$ and 2$d$, alternate around each hexagon.}
\label{fig8}
\end{figure}

In order to discriminate between the different possible structures, group theory and representation analysis were used to determine the irreducible representations compatible with the magnetic structure using a propagation vector of (1/2,~0,~0) via the BasIreps program within the FullProf suite \cite{Rodriguez1993}. Four possible one-dimensional irreducible representations are found, whose symmetry operators are listed in Table~\ref{TableIrreps}. They were tested by refining the differential powder diffractograms (2 - 35~K) with the scaling factor determined from the nuclear structure refinement. The fitting procedure was performed using a customized background in order to account for the step-like feature, which is actually an intrinsic signature of low-dimensional magnetic correlations as detailed below. Moreover a different Lorentzian broadening shifts with respect to the instrumental resolution width was implemented for the first two reflections in order to reproduce their distinct shape. A good refinement of the measured diffractogram difference could then be obtained only for the irreducible representation IRep(2) (see Fig.~\ref{fig7}). The resulting magnetic structure corresponds to two shifted honeycomb layers, each displaying antiferromagnetically coupled zig-zag ferromagnetic chains running along the $b$-axis (see Fig.~\ref{fig8}). In this magnetic arrangement, the main component of the magnetic moment is along the $b$-axis. Although allowed by symmetry, the refinement is not significantly improved by adding a component along $c$. Thus restricting the full magnetic moments along the $b$-axis, its amplitude is found equal to 2.95(3) and 2.5(3)~$\mu_B$ for the $2b$ and $2d$ Wyckoff sites respectively. Although this Rietveld analysis allows us to identify a zig-zag magnetic arrangement in this compound, it is only a first attempt since the low-dimensionality of the magnetic structure is not fully taken into account. In the next section, a more accurate approach based on Monte Carlo calculation is presented.

This magnetic structure corresponds to the C2'2'2$_1$ magnetic space group \cite{bilbaoserver,Schmid2008}, which forbids ferroelectricity. However, a linear magnetoelectric effect is allowed with a non-diagonal tensor of the form \cite{InternTable}:
\begin{equation*}
\begin{pmatrix}0&\alpha^{ME}_{12}&0\\ \alpha^{ME}_{21}&0&0\\ 0&0&0\end{pmatrix}
\end{equation*}
This tensor can be decomposed into a symmetric and an antisymmetric part. This compound is thus potentially ferrotoroidic as the toroidic moment is proportional to the antisymmetric part of the magnetoelectric tensor, hence allowed only when this tensor has non-diagonal terms \cite{Gorbatsevich1983}. A non-diagonal tensor was recently established in another honeycomb material, MnPS$_3$, however displaying another kind of collinear antiferromagnetic structure \cite{Ressouche2010}. 

The zig-zag magnetic structure of Na$_2$Co$_2$TeO$_6$ is one of the characteristic spin arrangements observed in honeycomb magnets. It can arise from Heisenberg interactions beyond the first neighbors in the classical and quantum $J_1$-$J_2$-$J_3$ phase diagram for a wide range of exchange parameters. It is also one of the phases stabilized in presence of strongly anisotropic interactions in the vicinity of the Kitaev spin liquid. It has already been evidenced in various materials in presence or not of strong spin-orbit coupling: e.g. Na$_2$IrO$_3$ \cite{Ye2012,Choi2012} and $\alpha$-RuCl$_3$ \cite{Sears2015,Johnson2015,Cao2016} for the former case, and BaNi$_2$(AsO$_4$)$_2$ \cite{Regnault1980} and $M$PS$_3$ with $M$=Ni, Co, Fe \cite{Kurosawa1983,Brec1986,Joy1992,Wildes2015} for the latter case. Closer to the present study, this magnetic arrangement has also been experimentally determined in Na$_3$Co$_2$SbO$_6$ \cite{Wong2016} and Na$_3$Ni$_2$BiO$_6$ \cite{Seibel2013}, and proposed to be stabilized in (Li,Na)$_3$Ni$_2$SbO$_6$ from ab-initio calculations \cite{Zvereva2015b}. Among these realizations, the peculiarity of Na$_2$Co$_2$TeO$_6$ is that it exhibits signature of both short-range and long-range magnetic correlations. This could be the consequence of some instability due to the vicinity of a phase boundary with a disordered phase or to remaining frustration effects. 
We have investigated the relevance of the $J_1$-$J_2$-$J_3$ isotropic Heisenberg model to our system using classical Monte-Carlo calculations. The deviations of our experimental observations from this classical model are expected to point out additional quantum effects and/or the relevance of Kitaev physics \cite{Khaliullin2005}.


\section{Monte Carlo calculations and discussion}

\begin{figure}
\resizebox{8cm}{!}{\includegraphics{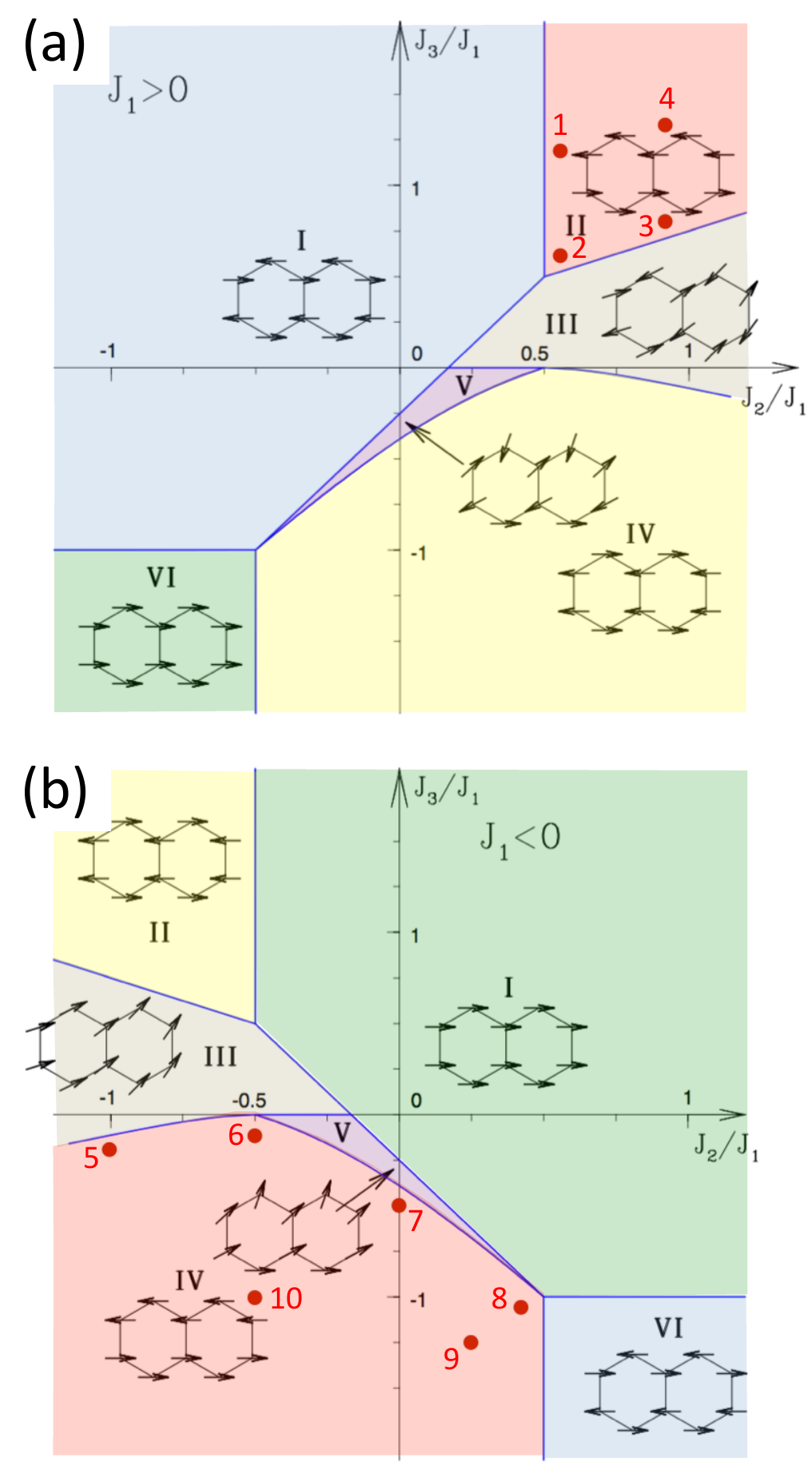}}
\caption{$J_2$/$J_1$--$J_3$/$J_1$ phase diagrams for AF $J_1>0$ (a) and for FM $J_1<0$ (b) (adapted from ref. \onlinecite{Fouet2001}). The related phases in both diagrams are similarly colored. The zig-zag ordered phase is in red and labelled II for AF $J_1$ and IV for FM $J_1$. The numbered red dots show the parameter sets that have been inspected through Monte-Carlo calculations.}
\label{fig9}
\end{figure}

To simulate the powder scattering function as well as thermodynamical quantities, an hybrid Monte Carlo method with a single-spin-flip Metropolis algorithm has been used on samples of $4\times L^3$ spins with $L$=12, 24. This algorithm is generally less efficient at low temperature because the number of rejected attempts increases with the development of spin correlations. To partially overcome this effect, the solid angle, from which each spin-flip trial is taken, is reduced to ensure that the acceptance rate is above 0.4 at every temperature. The scattering function (resp. thermodynamical quantities), has been averaged over 500 (resp. 10000) spin configurations at each temperature, while the number of Monte Carlo steps needed for decorrelation is adapted in such a way that the stochastic correlation between spin configurations is lower than 0.1. To improve this decorrelation and probe the configuration space more efficiently, a combination of overrelaxation \cite{Creutz1987} and molecular dynamics methods \cite{Taillefumier2014} has also been used.

The honeycomb lattice is not frustrated for AF nearest-neighbor isotropic interactions only, but frustration effects may arise when further neighbor interactions are present. Including these up to the third neighbors yields the $J_1$-$J_2$-$J_3$ model given by the following Hamiltonian, 
\begin{equation}
\mathcal{H}_{J_1J_2J_3}=  J_1\sum _{\langle  ij\rangle_1}{{\bf S}_{i} \cdot {\bf S}_{j}}+J_2\sum _{\langle  ij\rangle_2} {{\bf S}_{i} \cdot {\bf S}_{j}}+J_3\sum _{\langle  ij\rangle_3}{{\bf S}_{i} \cdot {\bf S}_{j}}
\label{eq1}
\end{equation}
The quantum and classical phase diagrams for this model have been established previously \cite{Rastelli1979,Fouet2001,Reuther2011,Albuquerque2011,Oitmaa2011,Li2016a,Li2016b} and intensively studied. The classical phase diagram, shown in Fig.~\ref{fig9} displays several ground states: ferromagnetic (FM, in green), two spiral (purple and grey) and several collinear AF (red, blue and yellow), among which the zig-zag one (red). There is a mapping between the $J_2$/$J_1$--$J_3$/$J_1$ phase diagrams with $J_1>0$ (AF) and $J_1<0$ (FM) \cite{Fouet2001}. The specific zig-zag magnetic structure is actually obtained either for FM or AF $J_1$, and exhibits some degeneracy of non-planar ground states that can be lifted by thermal/quantum fluctuations favoring the collinear solutions \cite{Fouet2001}. For AF $J_1$, $J_2$ and $J_3$ must be AF (see Fig.~\ref{fig9}(a)). In this case, the zig-zag phase is connected to a tricritical point of maximum degeneracy which was theoretically shown to host a classical spin liquid state \cite{Rehn2015}. When quantum fluctuations are included, this tricritical point smears out while one of the classical spiral phase (purple V) extends up to the zig-zag phase boundary, and becomes a quantum nonmagnetic phase (either plaquette valence bond crystal or spin liquid) \cite{Albuquerque2011,Reuther2011,Bishop2015}. In the FM $J_1$ case, $J_3$ must be AF and $J_2$ can be both AF and FM (see Fig.~\ref{fig9}(b)). No indication of any magnetically disordered phase was reported \cite{Li2012}.

To account for the observed three-dimensional ordering of Na$_2$Co$_2$TeO$_6$, a fourth AF interaction, $J_4>0$, linking the two honeycomb layers via the stacked Co$^{2+}$ ions at (0, 0, $\pm{ 1\over 4}$), has been added to the $J_1$-$J_2$-$J_3$ model. However, this interaction has been chosen much weaker than the reference interaction $|J_1|$, in order to reproduce the shape of the magnetic Bragg peaks which suggests less extended magnetic correlations perpendicular to the planes than within the planes. This is also suggested by the distant and indirect super-exchange paths linking two interacting atoms on adjacent layers. 

\begin{figure}
\resizebox{8.6cm}{!}{\includegraphics{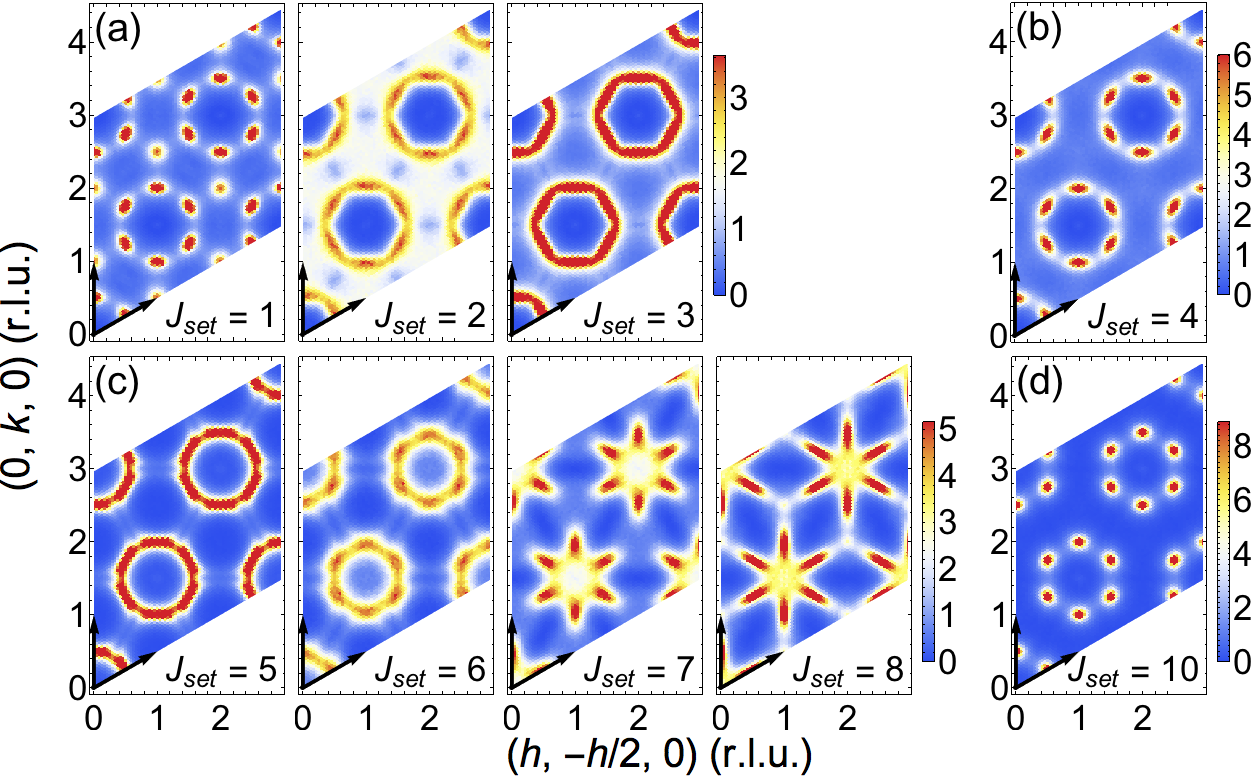}}
\caption{Calculated magnetic neutron scattering in the ($h$, $k$, 0) scattering plane for different positions in the $J_1>0$ (top) and $J_1<0$ (bottom) phase diagrams (labeled as in Fig.~\ref{fig9}) with $|J_4/J_1|$=0.025 and with reduced temperature t=T/$|J_1|>T_N$, equal to 0.5 (a and c) and equal to 1 (b and d).}
\label{fig10}
\end{figure}

\begin{figure}
\resizebox{8.0cm}{!}{\includegraphics{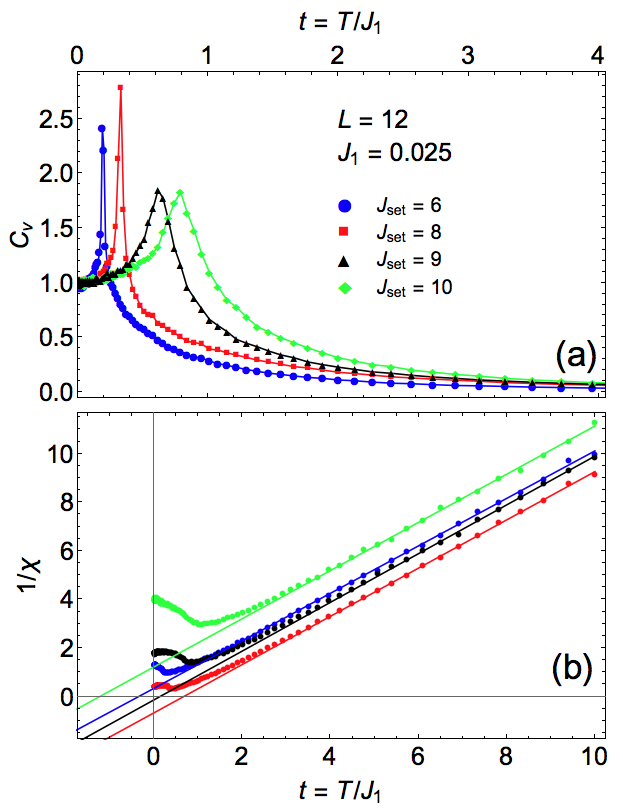}}
\caption{Calculated specific heat (a) and inverse susceptibility (b) for some of the investigated $J_{\rm set}$ in the $J_1<0$ phase diagram (labeled as in Fig.~\ref{fig9}) with $|J_4/J_1|$=0.025. The straight lines on the calculated $1/\chi$ is a Curie-Weiss fit yielding the Curie-Weiss temperature $\theta$ at the zero abscissa axis intercept.}
\label{fig11}
\end{figure}

\begin{figure}
\resizebox{8.6cm}{!}{\includegraphics{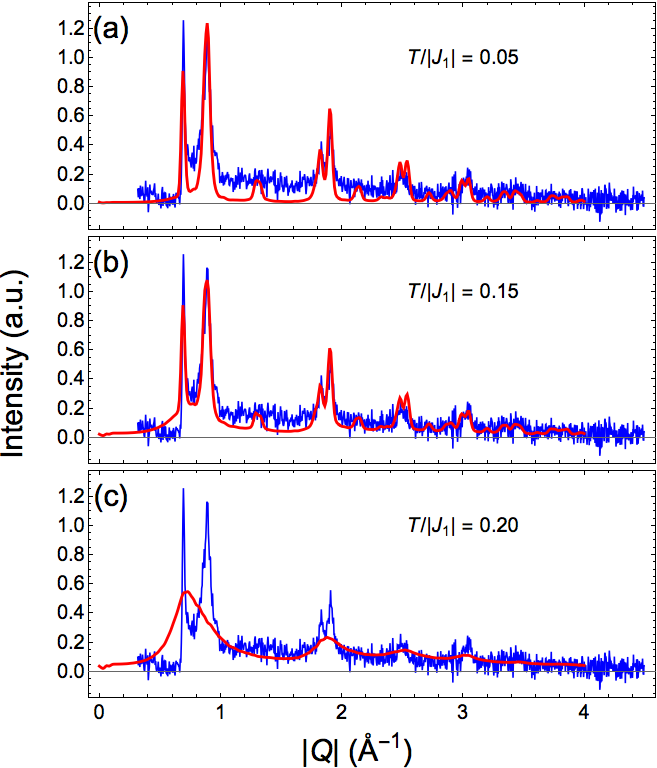}}
\caption{Calculated powder averaged magnetic neutron scattering (red line) versus measurements (blue line) for $J_2$/$J_1$=-0.5 and $J_3$/$J_1$=-0.1 ($J_{\rm set}$=6 in the $J_1<0$ phase diagram of Fig.~\ref{fig9}), and $|J_4/J_1|$=0.025 versus reduced temperature: t=T/$|J_1|$=0.05 (a), 0.15 (b), 0.20 (c). The experimental curve has been corrected from the subtracted paramagnetic scattering shown Fig. \ref{fig7}.}
\label{fig12}
\end{figure}

Starting from the simple Hamiltonian: 
\begin{equation}
\mathcal{H}= \mathcal{H}_{J_1J_2J_3}+J_4\sum _{\langle  ij\rangle_4}{{\bf S}_{i} \cdot {\bf S}_{j}}
\label{eq2}
\end{equation}
and assuming isotropic spins, we have calculated thermodynamics quantities such as specific heat and magnetic susceptibility and the neutron magnetic scattering versus the scattering vector $Q$ for various sets of $J_2/J_1$, $J_3/J_1$ and $J_4/J_1$ with positive or negative $J_1$ and for several reduced temperatures t=T/$|J_1|$. The investigated sets of $J_2/J_1$, $J_3/J_1$ parameters ($J_{\rm set}$) are shown in Fig.~\ref{fig9}. They were chosen both deep inside the zig-zag phase and at the proximity of the boundaries of this phase with adjacent ones. Note that the calculated neutron scattering was convoluted by a resolution function but also incorporates size effects in the powder averaging due to the finite size of the lattice. These yielded a global Gaussian-like resolution function (equal to 0.33 \AA$^{-1}$), which reproduces the experimental data rather well and is the same in all the calculations. 

A first result of the calculations concerns the magnetic diffuse scattering present above $T_N$, which reflects the magnetic correlations and varies significantly between the different parameter sets, as shown in the ($h$,$k$,0) plane in Fig.~\ref{fig10}. This is due in particular to the proximity of other phases with different ordering tendencies. However, below $T_N$, this diffuse scattering is masked by the rise of intense magnetic Bragg peaks. Moreover, the powder averaging, performed for comparison with the experimental data, blurs the differences. Finally, very similar powder structure factors are obtained which cannot be discriminated in the light of the measured one given the experimental uncertainties. 

Another noticeable difference between the behavior of the system when it is close or not from the phase boundaries is the temperature of magnetic ordering. This was roughly determined by locating the maximum of the calculated specific heat for $L$=12. It was checked not to vary significantly for $L$=24, although a full finite size scaling procedure should be employed for a more accurate determination. As seen in Fig.~\ref{fig11}, the transition temperature is systematically reduced when approaching a phase boundary, as an expected consequence of frustration due to competing orderings. However, the shape of the magnetic structure factor for identical values of T/$T_N$ in the ordered phase is very similar whatever the $J_1$-$J_2$-$J_3$ parameter set chosen in the zig-zag phase. 

The generic temperature evolution of the powder structure factor is presented in Fig.~\ref{fig12} for $J_{\rm set}$=6 ($J_1<0$ case) with small $|J_4/J_1|$=0.025. The distinct shapes of the first two peaks (thin for the first one and larger for the second one) is well accounted for in this model and is due to the small $J_4$ value as detailed afterwards. At the lowest temperature, well below $T_N/|J_1|\approx0.19$, the baseline is too low compared to the measured one. The step-like feature is recovered when approaching $T_N$ at the cost of all the peaks' sharpness. The general features of the experimental structure factor are thus rather well captured for a small $J_4$ value at temperatures below but close to T$_N$. Note however that the first narrow peak remains always too low in amplitude, which might be a consequence of additional in-plane disorder.

\begin{figure}
\resizebox{8.6cm}{!}{\includegraphics{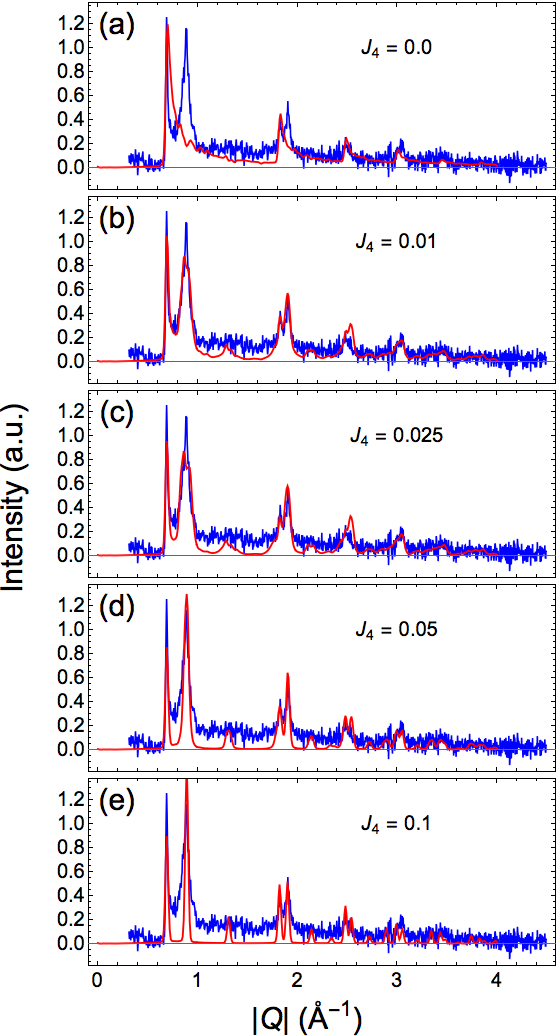}}
\caption{Calculated powder averaged magnetic neutron scattering (red line) versus measurements (blue line) for $J_2/J_1=-0.5$,  $J_3/J_1=-1$ ($J_{\rm set}$=10 in the $J_1<0$ phase diagram of Fig. \ref{fig9}) at t=T/$|J_1|$=0.15 for various values of $J_4/J_1$=0 (a), 0.01 (b), 0.025 (c) 0.05 (d) and 0.1 (e). The experimental curve has been corrected from the subtracted paramagnetic scattering shown Fig. \ref{fig7}.}
\label{fig13}
\end{figure}

The effect of the strength of the AF inter-layer coupling was also investigated. For $J_2/J_1=-0.5$ and $J_3/J_1=-1$ for instance ($J_{\rm set}$=10 in Fig.~\ref{fig9}), the evolution of the structure factors with different values of $J_4$ is shown in Fig.~\ref{fig13}. When $J_4=0$, a structure factor is calculated with asymmetric peaks and a step-like background, characteristics of the 2-dimensional ordering \cite{Warren41}. When increasing $J_4$, additional Bragg reflections (in particular the ($\pm 1/2$, 0, 1) one) start to rise. At the same time, the peaks characteristics of the in-plane ordering gets thiner and more symmetric, due to the establishment of out-of-plane magnetic correlations, and the step-like feature vanishes. 

Finally, some insight into the sign of the exchange interactions can be further obtained. An indication is given by the Curie-Weiss temperature which writes: 
\begin{equation}
\theta=-{S(S+1)\over k_B}(J_1+2J_2+J_3+J_4/3)
\label{eq3}
\end{equation}

We checked that, for the different $J_{\rm set}$ in the $J_1<0$ and $J_1>0$ phase diagram, the Curie-Weiss temperature numerically determined from the calculated inverse susceptibility was compatible with this analytical expression (see Fig.~\ref{fig11}). In the $J_1>0$ case, the calculated Curie-Weiss temperatures are strongly negative and much larger in absolute value than the $T_N$ temperature. This is at variance with the value determined from the measured susceptibility which is negative, thus pointing out dominant AF interactions, but smaller than the N\'eel temperature in absolute value. It strongly suggests competing AF and FM interactions in this system, in agreement with the $J_1<0$ phase diagram, on its $J_2>0$ side to account for the negative Curie-Weiss temperature. This is corroborated by the calculated Curie-Weiss temperatures which go from weakly negative (on the $J_2>0$ side) to weakly positive (on the $J_2<0$ side). Inspecting the super-exchange paths through the oxygen ions, it is found that the Co-O-Co angle for nearest neighbor Co is 92.175$^{\circ}$, which is compatible with a ferromagnetic $J_1$ according to the Goodenough-Kanamori rules. It should be noted however that an analysis including quantum fluctuations can slightly shift the magnetic susceptibility curve and will be necessary to go beyond the classical model \cite{Lohmann2014}.

To summarize, our Monte Carlo calculations have succeeded in reproducing our experimental neutron diffractogram using a $J_1$-$J_2$-$J_3$ model, most probably with FM $J_1$ and AF $J_2$ and $J_3$, supplemented by a weak AF $J_4$ interaction. This produces a system of weakly coupled honeycomb layers. It should be noted however that experimentally, the diffractograms recorded at 2 and 12~K (i.e. T/$T_N$=0.074 and 0.44 respectively) are almost identical. In the classical frame, no matter how small is $J_4$, the system should rapidly loose its 2-dimensional diffuse scattering signature and converge towards a well ordered phase with narrow and symmetric peaks and no step-like feature, at variance with the observation. The persistence well below $T_N$ of out-of-plane short-range correlations and low dimensional magnetism is therefore very intriguing. An attractive possibility is that quantum effects, not taken into account in the present calculations and that could be strongly enhanced at the proximity of a phase boundary, could stabilize this phase. The $J_{\rm set}$=6 for instance, close to the phase boundary with the spiral phase V and with medium AF $J_2$ interaction, fulfills those conditions. Another possibility is based on a distribution of inter-layer $J_4$ interaction. The resulting structure factor would thus be an average of the most correlated and less correlated ones calculated with different $J_4$ values, from zero to weakly antiferromagnetic. Since the interlayer Na atoms might participate, with the oxygens, to the magnetic exchange paths, a plausible explanation for this $J_4$ distribution relies on the strong Na disorder, intrinsically present in this material. However, preliminary calculations performed with a random distribution of $J_4$ values (not shown), do not seem to agree with the experimental structure factor. Therefore, for this scenario to be realized, an interlayer $J_4$ ordering must be present, for instance through a Na substructure \cite{Julien2008}, although it has not been experimentally evidenced yet in Na$_2$Co$_2$TeO$_6$.
 
While our model nicely captures some of the most intriguing magnetic behaviors of Na$_2$Co$_2$TeO$_6$, in particular the coexistence of robust short-range and long-range spin correlations, it does not explain other results revealed by macroscopic measurements such as the field-cooled/zero-field-cooled bifurcation at $T_N$, the two additional anomalies observed below $T_N$ and the metamagnetic process in the $M(H)$ in the ordered phase. Obviously, some ingredients are missing in our model, among which the single-ion magnetocrystalline anisotropy of the Co$^{2+}$ and the presence of two distinct crystallographic Co sites in the structure. To go further, in particular to determine qualitatively the Hamiltonian and to test the proximity of other phases as well as the signature of quantum effects, it is essential to measure the diffuse scattering and the magnetic excitation spectrum by neutron scattering on a single-crystal. This should allow to firmly establish whether or not the peculiar magnetic structure factor of Na$_2$Co$_2$TeO$_6$, combining short- and long-range low-dimensional magnetism, is a signature of the proximity of quantum phase boundaries. It should be noted that bond-directionality of the interactions, at the origin of Kitaev spin liquid on the honeycomb lattice, is also expected in cobaltates \cite{Khaliullin2005} and would be an interesting perspective to test.


\section{Conclusion}

In conclusion, we have disclosed the magnetic and magnetoelectric properties of Na$_2$Co$_2$TeO$_6$, a new honeycomb magnetic oxide. We have found that this compound stabilizes an antiferromagnetic zig-zag arrangement, one of the hallmark magnetic order in honeycomb magnets. It is associated to a non-diagonal magnetoelectric tensor compatible with ferrotoroidicity, a property that could be worth testing through future magnetoelectric measurements. This magnetic arrangement is obtained within the classical $J_1$-$J_2$-$J_3$ Heisenberg model with a weak inter-plane $J_4$ interaction, featuring a quasi 2-dimensional magnetism. However the temperature robustness of the magnetic structure factor, hence of the incomplete ordering, must imply additional quantum effects or interlayer coupling inhomogeneity. 


\begin{acknowledgments}
We would like to acknowledge J. Balay for his contribution in the preparation of the samples. We also thank S. Capelli for her help during the experiment on D1B. We are grateful to G. Khaliullin for interesting discussions on the Kitaev physics. This work was financially supported by Grant No. ANR-13-BS04-0013-01. 
\end{acknowledgments}


\begin{references}

\bibitem{Kitaev2006} A. Kitaev, Ann. Phys. {\bf 321}, 2 (2006).
\bibitem{Jackeli2009} G. Jackeli, G. Khaliullin, Phys. Rev. Lett. {\bf 102}, 017205 (2009).
\bibitem{Rastelli1979} E. Rastelli, A. Tassi, L. Reatto, Physica B {\bf 97}, 1 (1979)
\bibitem{Fouet2001} J. B. Fouet, P. Sindzingre, C. Lhuillier, Eur. Phys. J. B. {\bf 20}, 241 (2001).
\bibitem{Albuquerque2011} A. F. Albuquerque, D. Schwandt, B. Het\'enyi, S. Capponi, M. Mambrini, and A. M. L\"auchli, Phys. Rev. B {\bf 84}, 024406 (2011).
\bibitem{Reuther2011} J. Reuther, D. A. Abanin, R. Thomale, Phys. Rev. B {\bf 84}, 014417 (2011).
\bibitem{Oitmaa2011} J. Oitmaa, R. R. P. Singh, Phys. Rev. B {\bf 84}, 094424 (2011).
\bibitem{Li2016a} P. H. Y. Li, R. F. Bishop, C. E. Campbell, J. Phys.: Conf. Series {\bf 702}, 012001 (2016).
\bibitem{Li2016b} P. H. Y. Li, R. F. Bishop, Phys. Rev. B {\bf 93}, 214438 (2016).
\bibitem{Rehn2015} J. Rehn, A. Sen, K. Damle, R. Moessner, Phys. Rev. Lett. {\bf 117}, 167201 (2016).
\bibitem{Meng2010} Z. Y. Meng, T. C. Lang, S. Wessel, F. F. Assaad, A. Muramatsu, Nature {\bf 464}, 847 (2010). 
\bibitem{Viciu2007} L. Viciu, Q. Huang, E. Morosan, H.W. Zandbergen, N.I. Greenbaum, T. McQueen, R.J. Cava, Journal of Solid State Chemistry {\bf 180}, 1060 (2007).
\bibitem{Skakle1997} J. Skakle, S. T. Tovar, A. West, J. Solid State Chem. {\bf 131}, 115 (1997).
\bibitem{Zvereva2012} E.A. Zvereva, M.A. Evstigneeva, V.B. Nalbandyan, O.A. Savelieva, S.A. Ibragimov, O.S. Volkova, L.I. Medvedeva, A.N. Vasiliev, R. Klingeler, B. Buechner, Dalton Trans. {\bf 41} 572?580 (2012).
\bibitem{Zvereva2015a}  E.A. Zvereva, V.B. Nalbandyan, M.A. Evstigneeva, H.-J. Koo, M.-H. Whangbo, A. V. Ushakov, B.S. Medvedev, L.I. Medvedeva, N.A. Gridina, G.E. Yalovega, J. Solid State Chem. {\bf 225}, 89 (2015).
\bibitem{Zvereva2015b} E. A. Zvereva, M. I. Stratan, Y. A. Ovchenkov, V. B. Nalbandyan, J.-Y. Lin, E. L. Vavilova, M. F. Iakovleva, M. Abdel-Hafiez, A. V. Silhanek, X.-J. Chen, A. Stroppa, S. Picozzi, H. O. Jeschke, R. Valent\'i, and A. N. Vasiliev, Phys. Rev. B {\bf 92}, 144401 (2015).
\bibitem{Miura2006} Y. Miura, R. Hirai, J. Phys. Soc. Japan {\bf 75}, 084707 (2006).
\bibitem{Schmidt2013} W. Schmidt, R. Berthelot, A. Sleight, M. Subramanian, J. Solid State Chem. {\bf 201}, 178 (2013).
\bibitem{Berthelot2012a} R. Berthelot, W. Schmidt, A. Sleight, M. Subramanian, J. Solid State Chem. {\bf 196}, 225 (2012).
\bibitem{Sankar2014} R. Sankar, I.P. Muthuselvam, G. Shu, W. Chen, S.K. Karna, R. Jayavel, F. Chou, CrystEngComm {\bf 16} 10791 (2014).
\bibitem{Xu2005} J. Xu, A. Assoud, N. Soheilnia, S. Derakhshan, H.L. Cuthbert, J.E. Greedan, M. H. Whangbo, H. Kleinke, Inorg. Chem. {\bf 44}, 5042 (2005).
\bibitem{Berthelot2012b} R. Berthelot, W. Schmidt, S. Muir, J. Eilertsen, L. Etienne, A. Sleight, M. A. Subramanian, Inorg. Chem. {\bf 51}, 5377 (2012).
\bibitem{Seibel2013} E.M. Seibel, J. Roudebush, H. Wu, Q. Huang, M.N. Ali, H. Ji, R. Cava, Inorg. Chem. {\bf 52}, 13605 (2013).
\bibitem{Wong2016} C. Wong, M. Avdeev, C. D. Ling, J. Solid State Chem. {\bf 243}, 18 (2016).
\bibitem{Takada2003} K. Takada, H. Sakurai, E. Takayama-Muromachi, F. Izumi, R. A. Dilanian, T. Sasaki, Nature {\bf 422}, 53 (2003).
\bibitem{Kittel1976} C. Kittel, {\it Introduction in Solid State Physics}, 5th ed., Wiley, New York, 1976.  
\bibitem{Rodriguez1993} J. Rodriguez-Carjaval, Physica B {\bf 192}, 55 (1993).
\bibitem{bilbaoserver} J.M. Perez-Mato, S.V. Gallego, E.S. Tasci, L. Elcoro, G. de la Flor, and M.I. Aroyo, Annu. Rev. Mater. Res. {\bf 45}:13.1-13.32 (2015).
\bibitem{Schmid2008} H. Schmid, J. Phys.: Condens. Matter {\bf 20} (2008) 434201.
\bibitem{InternTable} A. S. Borovik-Romanov, H. Grimmer, {\it International Tables for Crystallography}, Vol. D, chapter 1.5, p138 (2006).
\bibitem{Gorbatsevich1983} A. A. Gorbatsevich, and Y. V. Kopaev, and V. V. Tugushev, Sov. Phys. JETP {\bf 58}  (1983) 643.  
\bibitem{Ressouche2010} E. Ressouche, M. Loire, V. Simonet, R. Ballou, A. Stunault, and A. Wildes, Phys. Rev. B {\bf 82}, 100408(R) (2010).
\bibitem{Ye2012} F. Ye, S. Chi, H. Cao, B. C. Chakoumakos, J. A. Fernandez-Baca, R. Custelcean, T. F. Qi, O. B. Korneta, G. Cao, Phys. Rev. B {\bf 85} 180403 (2012).
\bibitem{Choi2012} S. K. Choi, R. Coldea, A. N. Kolmogorov, T. Lancaster, I. I. Mazin, S. J. Blundell, P. G. Radaelli, Y. Singh, P. Gegenwart, K. R. Choi, S.-W. Cheong, P. J. Baker, C. Stock, and J. Taylor, Phys. Rev. Lett. {\bf 108}, 127204 (2012).
\bibitem{Johnson2015} R. D. Johnson, S. C. Williams, A. A. Haghighirad, J. Singleton, V. Zapf, P. Manuel, I. I. Mazin, Y. Li, H. O. Jeschke, R. Valent\'i, and R. Coldea, Phys. Rev. B {\bf 92}, 235119 (2015).
\bibitem{Sears2015} J. A. Sears, M. Songvilay, K. W. Plumb, J. P. Clancy, Y. Qiu, Y. Zhao, D. Parshall, Y.-J. Kim, Phys. Rev. B {\bf 91} 144420 (2015).
\bibitem{Cao2016} H. B. Cao, A. Banerjee, J.-Q. Yan, C. A. Bridges, M. D. Lumsden, D. G. Mandrus, D. A. Tennant, B. C. Chakoumakos, S. E. Nagler, Phys. Rev. B {\bf 93}, 134423 (2016).
\bibitem{Regnault1980} L. P. Regnault, J. Y. Henry, J. Rossat-Mignod, and A. deCombarieu, J. Magn. Magn. Mater. {\bf 15}, 1021 (1980).
\bibitem{Kurosawa1983} K. Kurosawa, S. Saito, and Y. Yamaguchi, J. Phys. Soc. Japan {\bf 52}, 3919 (1983).
\bibitem{Brec1986} R. Brec, Solid State Ionics {\bf 22}, 3 (1986).
\bibitem{Joy1992} P. A. Joy and S. Vasudevan, Phys. Rev. B {\bf 46}, 5425 (1992).
\bibitem{Wildes2015} A. R. Wildes, V. Simonet, E. Ressouche, G.J. McIntyre, M. Avdeev, E. Suard, S. A. J. Kimber, D. Lan\c{c}on, G. Pepe, B. Moubaraki, and T. J. Hicks, Phys. Rev. B {\bf 92}, 224408 (2015).
\bibitem{Khaliullin2005} G. Khaliullin, Prog. Theor. Physics Suppl. {\bf 160}, 155 (2005).
\bibitem{Creutz1987} M. Creutz, Phys. Rev. D {\bf 36}, 515 (1987).
\bibitem{Taillefumier2014} M. Taillefumier, J. Robert, C. L. Henley, R. Moessner, B. Canals, Phys. Rev. B {\bf 90}, 064419 (2014).
\bibitem{Bishop2015} R. F. Bishop, P. H. Y. Li, O. G\"otze, J. Richter, C. E. Campbell, Phys. Rev. {\bf B 92}, 224434 (2015).
\bibitem{Li2012} P. H. Y. Li, R. F. Bishop, D. J. J. Farnell, J. Richter, C. E. Campbell, Phys. Rev. B {\bf 85}, 085115 (2012).
\bibitem{Warren41} B. E. Warren, Phys. Rev. {\bf 59}, 693 (1941).
\bibitem{Lohmann2014} A. Lohmann, H.-J. Schmidt, J. Richter, Phys. Rev. B {\bf 89}, 014415 (2014).
\bibitem{Julien2008} M.-H. Julien, C. de Vaulx, H. Mayaffre, C. Berthier, M. Horvati\'c, V. Simonet, J. Wooldridge, G. Balakrishnan, M. R. Lees, D. P. Chen, C. T. Lin, P. Lejay, Phys. Rev. Lett. {\bf 100}, 096405 (2008).

\end{references}
\end{document}